**Free electron charging of microdroplets in a plasma at atmospheric pressure**

Nourhan Hendawy,[1,2,*] Harold McQuaid,[3,§] Somhairle Mag Uidhir,[3] David Rutherford,[4] Declan Diver,[5] Davide Mariotti,[6,†] Paul Maguire[5]

[1]School of Life Sciences, Sussex University, Brighton, United Kingdom
[2]Biaco Energy Ltd, Hassocks, United Kingdom
[3]School of Engineering, Ulster University, Belfast, N. Ireland
[4]Faculty of Electrical Engineering, Czech Technical University, Technická 2, Prague 6 Czech Republic
[5]School of Physics and Astronomy, University of Glasgow, Glasgow, Scotland
[6]Faculty of Engineering, University of Strathclyde, Glasgow, Scotland

**ABSTRACT**. Gas-phase microdroplets have recently demonstrated exceptional chemical properties via suspected mechanisms such as contact electrification and surface charge pinning, producing high surface electric fields. Here, microdroplets are injected into a low-temperature atmospheric-pressure plasma and exposed to an excess free-electron flux. We report the first measurements of droplet charge in a fully collisional plasma, with averages up to $2.5 \times 10^5$ electrons for 15 µm droplets, dependent on absorbed power. Simulations indicate solid particles under similar conditions acquire ~40% less charge, likely due to low-mobility water cluster ion formation around evaporating droplets, and predict surface electric fields up to $10^7$ V m$^{-1}$ for the smallest droplets (3 µm). Plasma exposure also produces $H_2O_2$ in the liquid, with concentrations up to 33 mM, corresponding to remarkable generation rates exceeding 275 M s$^{-1}$. The system involves complex, not yet fully elucidated mechanisms. A controlled plasma environment enables defined charge levels and exposure times, allowing systematic study and enhancing droplet properties.

**I. INTRODUCTION**.

Gas-phase microscale liquid droplets have a pervasive presence and impact in nature including cloud formation, atmospheric chemistry and geological erosion with possible relevance to prebiotic origin theories, where compartmentalisation may have had a significant effect on reaction thermodynamics[1–5]. The unique reaction environment provided by microdroplets is demonstrating an ever increasing significance in many scientific and technological fields from materials and drug synthesis to biological medicine[6–11]. Droplet-based microreactors have also recently generated widespread interest for precision chemical synthesis due to reaction rate enhancement of many orders of magnitude[3,6,12,13]. Initial explorations have shown the ability with microdroplets to overcome reaction barriers that are extremely challenging in bulk liquids with important climate-related consequences for high-energy industrial processes such as ammonia production and $CO_2$ transformation[6,8,14–17]. Microdroplets have also been found to spontaneously induce redox reactions with implications for a wide range of chemical processes including those in living systems, such as photosynthesis and respiration[5,18,19].

Improved control of thermodynamic parameters, reaction kinetics, heat and mass transfer and chemical mixing are expected advantages of moving from the bulk liquid to the microscale. The high surface to volume ratio, the effects of partial solvation on reaction rates are potentially important as well as the suspected presence of surface charge and electric field which can give rise to ultrathin electronic double layers; ion separation, confinement and alignment; enhanced concentration gradients and diffusive flows; solvent dipole alignment and reduced permittivity; extreme pH conditions, charge transport and spontaneous redox reactions, induced ionisation of gas and vapour molecules, droplet fission and plasma formation[2,10,13]. Charge levels and field intensities are unknown with various alternative electrification and charge transfer mechanisms suggested[2,3,9,12,17,18,20,21].

Microdroplets represent a complex physical and chemical system with multiple interconnected and transient parameters that are not readily amenable to experimental control and poses considerable challenges for detailed characterization. Observed effects are critically dependent on size, with an upper limit of approximately 15 µm diameter, and a lower limit of a few microns, determined by lifetime and technical limits. Acquired surface charge is subject to recombination with gas-phase ions during flight. Droplet size and hence internal conditions evolve during flight, due to evaporation, fission and collisions, limiting experimental exposure typically to the sub-millisecond range. Electric fields, concentration gradients, solvent alignment and charge transfer are sensitive to size through their dependence on surface curvature and surface area to volume ratio. Thus, the development of a comprehensive theoretical framework which is necessary for understanding the impact of microdroplets in nature and their exploitation in applications, is significantly hampered.

\* nourhan.hendawy@biaco.energy
§h.mcquaid@ulster.ac.uk
†davide.mariotti@strath.ac.uk

Microdroplet charging by plasma offers an alternative or complementary approach that could allow greater elucidation of fundamental characteristics. On entry into a low-temperature non-equilibrium plasma region, the high electron temperature ensures an initial net flux of high mobility free electrons from the gas phase to the droplet surface, creating a layer of surface charge and an associated electric field. After the droplet floating potential is established, over a few nanoseconds, the electron and positive ion flux are equalized. Thereafter the electrical characteristics remain reasonably constant over a known, and controllable, time period determined by plasma length and droplet velocity. A suitable plasma is required to operate at atmospheric pressure with low gas temperature to limit droplet evaporation and hence a high gas flow and restricted plasma size is inevitable, presenting a challenge for injection of microscale droplets and charge measurement.

There has been a long-standing interest in plasma charging of solid particles in the field of dusty (or complex) plasmas in order to study phenomena ranging from astrophysical (lunar exosphere, Saturn's rings, planetary debris) to plasma crystals, magnetic confinement fusion research, semiconductor processing and quantum dot synthesis. [22–24] However, charge measurements in high pressure dusty plasmas have been restricted to nano-sized particle assemblies, and have matched predictions of numerical simulations[25–27]. The charge on microscale particles, however, has been measured only at low pressure where ensemble levitation experiments are possible. Estimates of particle mean charge range from $10 - 10^4$ electrons for particles up to ~10 µm in diameter, which is similar to ion charging in unipolar corona environments[28–32].

In this paper, we generate a steady low density stream of microdroplets, with a mean diameter of 13 µm, injected into the plasma gas flow, with the gas temperature maintained in the range 20 ºC – 40 ºC, depending on input power[33,34]. The plasma volume contains an average of 6 microdroplets at any instant, therefore collisions are unlikely and the depletion of plasma electrons by the accumulation of surface charge is expected to be negligible. From the measured evaporation rate, > 95% are expected to survive the plasma flight time of ~ 120 µs. Microdroplets, unlike solid particles, may be surrounded by a vapour rich region that could influence the total positive ion flux due to the formation of low mobility positive water-cluster ions. The average microdroplet charge is determined from net current measured downstream of the plasma afterglow region and is compared with finite element simulations based on a fully hydrodynamic drift-diffusion model with a range of possible ion mobilities evaluated.

## II. METHODS

### Plasma details

A radio frequency (RF) driven (13.56 MHz) atmospheric pressure plasma was generated inside a 2 mm inner diameter quartz tube, between two exterior concentric Cu electrodes, ~2 mm apart and operated in helium, FIG 1. The plasma impedance and net power absorption were monitored continuously using an inline voltage-current (VI) probe (Impedans Octiv Suite 2.0). The water vapour content, monitored using an inline Xentaur XTR-100 Dew Point Meter, was maintained below 100 ppm, without droplets present[35]. The gas temperature was continuously monitored using a Micro-Epsilon CT precise infrared sensor (CT-CF22-C1 Miniature-Pyrometer)[36]. Maximum gas temperatures were 46 ºC for a net absorbed power of 0.6 W at the lowest flow rate (0.6 sLm) and without droplets. The introduction of microdroplets reduced the gas temperature by up to 10 ºC.

\* nourhan.hendawy@biaco.energy  
§h.mcquaid@ulster.ac.uk  
†davide.mariotti@strath.ac.uk

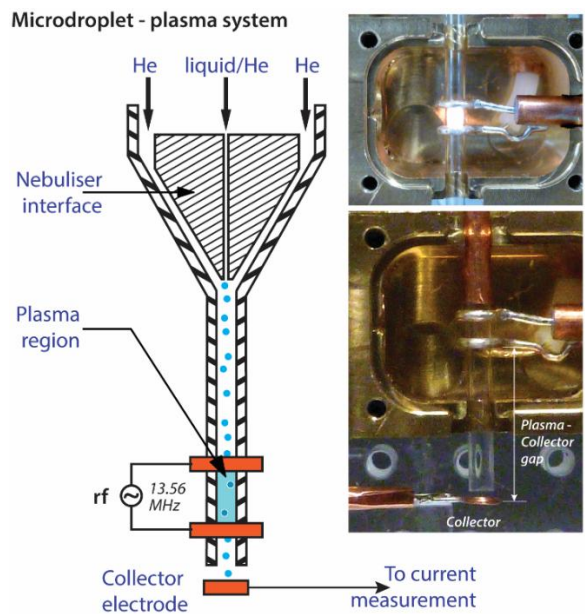

FIG 1: RF atmospheric pressure plasma (APP), microdroplets and collector electrode configuration.

**Droplet size and velocity distributions**

A continuous stream of near mono-sized microdroplets was obtained from a Burgener Mira Mist X-175 nebuliser with average diameters of 13 $\mu$m. The nebuliser was driven by a He gas flow ($Q_1$) via mass flow controller (MKS Type GE50A, PR4000B) while the liquid (H$_2$O) flow ($Q_L$) was supplied from a syringe pump. The divergent (20°) output of the nebuliser was interfaced to a quartz capillary tube with a 2 mm inner diameter via a custom-designed acrylic manifold, FIG 1, which focussed the droplet/He gas flow ($Q_1$) using a concentric helium gas curtain with flow values ($Q_2$) variable between 1 – 4 sLm, via a mass flow controller. The droplet transport efficiency was optimized by varying $Q_L$ (5 mL min$^{-1}$), $Q_1$ (0.7 sLm) and $Q_2$ (3.5 sLm). The total plasma gas flow was therefore $Q_{pl} = Q_1 + Q_2$. Droplet velocity, position and diameter (D) after exiting the plasma region were obtained from high resolution images of area 1805 x 1805 $\mu$m centred at 1500 $\mu$m and 4000 $\mu$m from the end of the capillary tube. A Questar QM-100 long focal length microscope was used to image the droplets onto an Andor IStar CCD camera with exposure times up to 20 $\mu$s. Velocity and diameter distributions were determined from streak length and width respectively from a sample of droplets (n > 1000). The path lengths from nebuliser to plasma region and to capillary exit were ≥ 50 mm and 60 mm – 70 mm respectively. Droplet trajectories imaged with and without plasma indicated laminar flow conditions within the capillary. Low resolution images with a larger field of view show a parallel droplet spray at the capillary tube outlet, indicating persistent laminar flow beyond the capillary, with a slight reduction in spray diameter when the plasma is turned on.

For absorbed plasma power of 1 W and total gas flow ($Q_1 + Q_2$) of 3.0 sLm, the droplet size statistics indicated a lognormal distribution with count median diameter = 13.6 mm and geometric std. deviation = 1.7, FIG 2. The total droplet rate was 5 x 10$^4$ s$^{-1}$. The average and peak gas velocities within the capillary are ≤ 16 ms$^{-1}$ and ≤ 32 ms$^{-1}$ respectively, assuming laminar flow conditions. The measured velocity distribution, outside the capillary, indicates droplet deceleration with 50% of droplets travelling at ≤ 12 ms$^{-1}$. Measurements were obtained at an average distance of 1.5 mm from the plasma capillary outlet.

* nourhan.hendawy@biaco.energy
§h.mcquaid@ulster.ac.uk
†davide.mariotti@strath.ac.uk

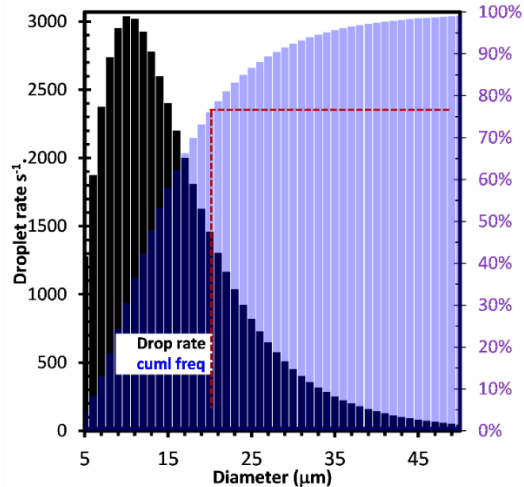

FIG 2: Droplet rate versus size. Also shown is the cumulative probability. The net plasma power was 1 W and the total gas flow was 3.0 sLm.

## III. RESULTS AND DISCUSSION

The measured current consists of direct plasma electron and ion flux components as well as a component due to the charge on the droplets. The latter is extracted by positioning the collector beyond the electron recombination limit, estimated from the length of the luminous plasma afterglow region, while the remaining ion current, $i_{ion}$, is estimated from plasma measurements without droplets. In FIG 3, the collector potential versus power is shown for plasmas with and without droplets, with the collector electrode at 3 mm distance from the nearest plasma electrode. Without droplets, the collector potential is due to $i_{ion}$ only and increases with absorbed power until a point of inflection at 0.6 W. This indicates the power threshold at which the plasma afterglow makes electrical contact with the collector. With droplets present, the plasma afterglow length is reduced considerably and higher absorbed power values, up to 5 W, are possible. The linear fit to $i_{ion}$ is obtained from data up to the power threshold (shaded region) and the net potential versus power characteristic, $(i_{meas} - i_{ion})R_o$, is obtained by subtraction of the $i_{ion}$ fit from the linear fit to $i_{meas}$, and the result is attributed to the charge carried by the droplet stream, as discussed in the Supplemental Material[37]. A voltage to charge conversion factor of $1.1 \times 10^{-7}$ V electron$^{-1}$ was obtained from simulation, see Supplemental Material[38].

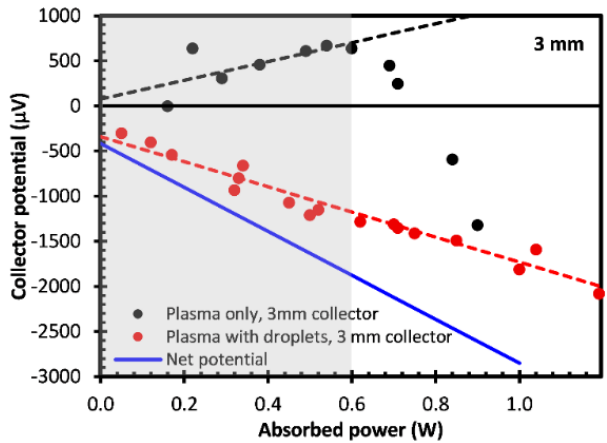

FIG 3. Collector potential (VC) versus absorbed power, with and without droplets at a distance of 3 mm, using a 3 mm diameter disk collector to measure droplet charge. Dashed lines represent least squares fit to measurements over the power range (shaded) where the plasma afterglow does not reach the collector. With droplets, the afterglow region is shortened and measurements extend to higher powers. The net potential due to droplets alone is obtained from subtracting the plasma fitted line (black) from the droplets fitted line (red).

* nourhan.hendawy@biaco.energy
§h.mcquaid@ulster.ac.uk
†davide.mariotti@strath.ac.uk

In FIG 4, the droplet charge versus collector distance is shown for low (1.5 W) and high (5 W) powers. The droplet charge in the plasma region (x = 0) was estimated from a sigmoidal fit to the charge – distance curve and the maximum value was $2.5 \times 10^5$ electrons (40 fC), approximately 2 orders of magnitude higher than that achievable by other approaches, including corona and in low pressure collisionless plasmas.

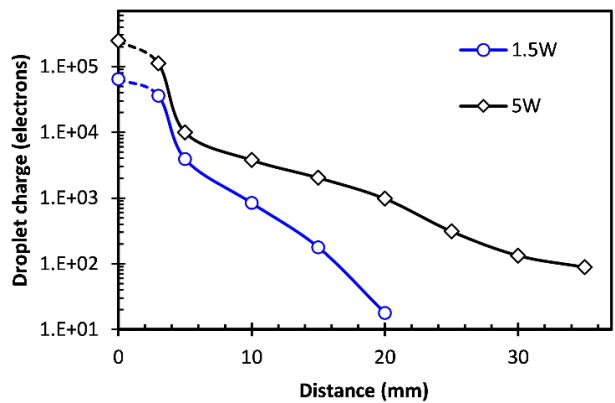

FIG 4. Droplet charge (electrons) versus distance from plasma. The data points are obtained from the subtraction of linear fits to potential versus power with and without droplets, at each distance, to give the droplet induced potential. The charge conversion factor was $1.1 \times 10^{-7}$ V electron$^{-1}$. The collector diameter was 3 mm.

These measurements of microdroplet charge represent the first such measurements of free electron microparticle charging in a fully collisional plasma. Simulation of ion and electron fluxes to the droplet, based on a hydrodynamic mobility model appropriate for a fully collisional plasma, and assuming $He_2^+$ as the dominant ion, resulted in predicted charge levels up to 50% lower than that measured (see Supplemental Material[39]). The value of plasma density, $10^{20}$ m$^{-3}$, was determined from in-situ plasma impedance measurements (see Supplemental Material[40]). while electron temperature, 2 eV, was estimated from plasma simulation whereby $T_e$ was adjusted to match the simulated OH$^\bullet$ and $H_2O_2$ fluxes to experimental measurements[35]. To account for the distribution of droplet sizes, and hence charge, the simulations were carried out for radius values between 3 μm and 25 μm and a charge – size relationship determined using a polynomial fit. The equivalent weighted average charge was then obtained by applying the fit to the measured distribution of droplet sizes.

The dynamic interface between the liquid and gaseous phases likely involves significant bidirectional charge and mass transfer leading to a vapour-rich region surrounding the droplet that would alter the ion flux constituents, impacting the ion mobility and hence the net charge. We previously determined the droplet evaporation rate to be about one order of magnitude higher than reported for droplet evaporation in a flowing gas[34]. Assuming diffusion of evaporated vapour molecules away from the droplet, under laminar flow conditions, this vapour-rich region would extend approximately 100 μm away from the droplet surface after the ~ 120 μs flight time in the plasma. This is significantly larger than the predicted sheath width around the droplet, which is typically of the order of the droplet radius itself. Normally, relatively low $H_2O$ concentrations (~ 1%) would extinguish a plasma[35]. However, with a high $H_2O$ concentration halo around the droplet, charge irradiation from the surrounding electropositive He plasma would promote a distinct halo plasma chemistry which would include $H_2O$ dissociation products (i.e. $H_iO_j$, i,j: 0 → 2) in neutral and ionic form. Multiple collisions between such ions and water molecules lead to the growth of large positive, $H^+(H_2O)_n$, and negative $OH^-(H_2O)_n$, water clusters ions, the cluster size increasing with vapour content, [41–43]. Large positive cluster ions with n up to 55 (~1000 amu) have been reported for a He plasma jet into air[44]. Were such large ions to constitute a significant fraction of the positive ion density in the droplet neighbourhood, then the ion flux to the droplet surface is reduced due to the much lower ion mobility, up to 2 orders of magnitude compared to $He_2^+$, and the surface negative charge is thereby increased[45]. Using an ion mobility value of $10^{-5}$ m$^2$ V$^{-1}$ s$^{-1}$, the simulated charge values, FIG 5, matched those measured, which implicates heavy cluster ions as the dominant constituent of the positive ion flux. However additional physical mechanisms may also need to be considered including electron reflection, secondary electron emission, surface recombination of $He_2^+$ ions and He metastables. Although current understanding of Auger processes in dielectric liquids is sparse, particularly for low energy ion irradiation, the possibility exists for Auger neutralisation leading to an enhancement of the electron flux over that of the ions by a factor given by the secondary electron emission coefficient[46–48]. Furthermore, any emitted hot electrons are then absorbed within a mean free path of the surface leading to processes that may include resonant ionisation of He metastables and water dissociation or ionisation which may increase the local plasma density.

* nourhan.hendawy@biaco.energy
§h.mcquaid@ulster.ac.uk
†davide.mariotti@strath.ac.uk

Also shown in FIG 5 is the variation in surface electric field with size. The maximum field, $1.1 \times 10^7$ V m$^{-1}$, is observed at the smallest radius, 3 µm. This is unlikely to be sufficient to cause gas dissociation or ionisation at the interface or electron stripping from surface hydroxide molecules to create OH• radicals, as previously suggested for microdroplets with levels of estimated charge much lower than reported here[21,49]. The alternative hypothesis, namely high transient electric fields occurring during droplet fission, would only be relevant when droplet size reduces to the Rayleigh limit for fission[2,9,21]. Measured charge levels are ~ 8% of the limit and fission is therefore not expected until the droplet size reaches ~1.5 µm, which is below the survival size limit due to evaporation within the plasma and hence will not contribute to the charge measurement. Chemical reaction rates within the microdroplet are enhanced, as with other microdroplet sources, due to surface charge and field effects leading to a surface double layer within the droplet and enhanced concentration gradients.

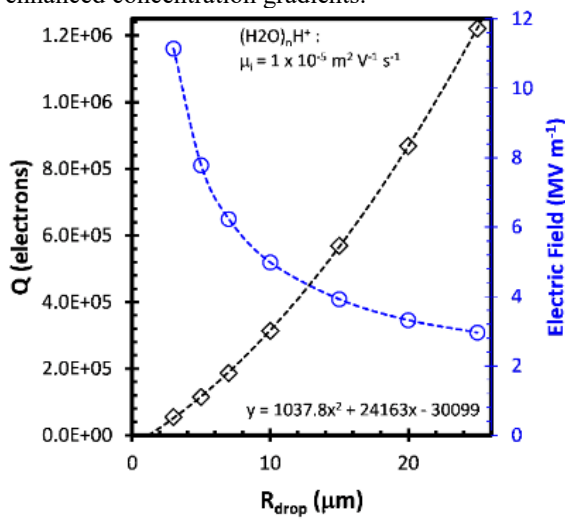

FIG 5. Charge and electric field versus droplet radius obtained from simulation ($\mu_{ion} = 10^{-5}$ m$^2$ V$^{-1}$ s$^{-1}$, $T_e = 2$ eV, $n_e = 10^{20}$ m$^{-3}$) and equivalent to a weighted average charge across the droplet size distribution of $2.5 \times 10^5$ electrons.

Free electron charging in a plasma also provides a direct source of reactive agents, namely electrons, with a constant replenishment flux over the plasma flight time. Flux estimates from simulation are $10^{20}$ m$^{-2}$ s$^{-1}$ to $10^{21}$ m$^{-2}$ s$^{-1}$ with increasing size. This places an upper limit on the electron-initiated reaction enhancement, across all possible reactions, of ~ 3 M s$^{-1}$ for the smallest droplets, decreasing to 26 mM s$^{-1}$ for the largest and supports our previous reports of metal ion reduction in microdroplets, where rate enhancements of $> 10^6$ over bulk liquid were observed[50]. We measured the concentration of $H_2O_2$ in plasma charged droplets at $> 33$ mM, after 120 µs flight time, equivalent to a remarkable generation rate of 275 M s$^{-1}$, even by comparison with the microdroplet rate enhancement recently reported[7,13,51]. The $H_2O_2$ generation rate due to plasma alone, without microdroplets, was $< 100$ nM s$^{-1}$. Spontaneous production of $H_2O_2$ has been observed in nominally uncharged microdroplets with concentrations up to 10 µM reported and attributed to sequestering of atmospheric gas impurities, e.g. $O_3$, or enhanced electric fields due to droplet fission induced by high gas flows. We estimate the plasma-irradiation of microdroplets leads to an enhancement of the $H_2O_2$ generation rate by $> 10^5$ compared to recent literature[18,49,52]. Estimates of charge values up to $4.5 \times 10^4$ electrons, in similarly produced droplet flows, have been reported[20], although a sensitivity to gas flow conditions has been observed [52]. While the charge levels are commensurate with those measured here (within a factor of 10), the extremely high generation rates we observe are likely due to the additional impact of solvation of free electrons arriving at the surface in tandem with gas species sequestering and enhanced concentration gradients due to the formation of a double layer within the liquid. These results provide initial insight into the dynamic and complex microdroplet – plasma system, driven by bidirectional mass and charge transfer, that exhibits unique and interesting properties, including significantly elevated chemical reaction rates, and deserving of further detailed study of the underlying physical mechanisms.

* nourhan.hendawy@biaco.energy
§h.mcquaid@ulster.ac.uk
†davide.mariotti@strath.ac.uk

## ACKNOWLEDGMENTS

This work was supported by Engineering and Physical Sciences Research Council (Project Nos. EP/K006088/1, EP/K006142/1, EP/K022237/1, EP/R008841/1, EP/T016000/1) and EU COST Actions PlAgri (CA19110) and PlasTHER (CA20114).

## AUTHOR CONTRIBUTIONS

The authors contributed equally to this work. Correspondence and requests for materials should be addressed to P.M. (email: paul.maguire@glasgow.ac.uk)

## DATA AVAILABILITY

The data that support the findings of this study are available from the authors on reasonable request.

* nourhan.hendawy@biaco.energy
§ h.mcquaid@ulster.ac.uk
† davide.mariotti@strath.ac.uk

* nourhan.hendawy@biaco.energy
§h.mcquaid@ulster.ac.uk
†davide.mariotti@strath.ac.uk

\* nourhan.hendawy@biaco.energy
§h.mcquaid@ulster.ac.uk
†davide.mariotti@strath.ac.uk

# Supplemental Material

# Free electron charging of microdroplets in a plasma at atmospheric pressure

Nourhan Hendawy[1,2*], Harold McQuaid[3§], Somhairle Mag Uidhir[3], David Rutherford[4], Declan Diver[5], Davide Mariotti[6†], Paul Maguire[5]

[1]School of Life Sciences, Sussex University, Brighton, United Kingdom
[2]Biaco Energy Ltd, Hassocks, United Kingdom
[3]School of Engineering, Ulster University, Belfast, N. Ireland
[4]Faculty of Electrical Engineering, Czech Technical University, Technická 2, Prague 6 Czech Republic
[5]School of Physics and Astronomy, University of Glasgow, Glasgow, Scotland
[6]Faculty of Engineering, University of Strathclyde, Glasgow, Scotland

## Section S1. Current measurements

We measure the average droplet charge using a low-density stream of microdroplets impacting a contact electrode after passing through a short (2 mm) atmospheric pressure plasma region. A disk electrode larger than the capillary diameter ($R_C \geq 1.5$ mm) was positioned downstream from the plasma with sufficient gap to ensure placement beyond the plasma afterglow region and hence minimise plasma interference. A small airgap of $\leq 1$ mm was also maintained between the capillary outlet and the electrode to minimise turbulence effects and possible droplet loss.

A charged particle approaching a conducting electrode creates an image charge and with a finite collector electrode geometry, the magnitude of the image charge, $q'$, is reduced and given approximately by[1]

$$i = \frac{dq'}{dt} = \frac{qvR_C^2}{\sqrt{(R_C^2 + x^2)^3}} \qquad (1)$$

where x is the distance between the droplet and collector electrode, $R_C$ is the collector radius and v the droplet velocity. As $x \to 0$, $i \to qv/R_C$. We tested the validity of (1) using single large droplets, ~ 2 mm in diameter, of known charge ($10^8$ electrons) and floating potential (150V), obtained from a DC biased hypodermic needle and allow to free fall onto a collector plate. The velocity and size were determined from imaging. Measured current pulses, Figure S1, matched the prediction from (1)

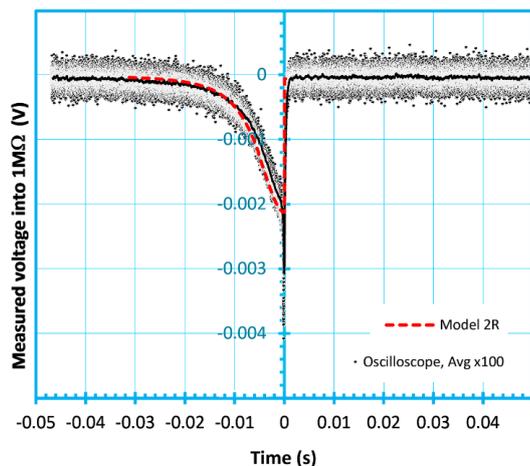

**Figure S1:** Current measurements for a 2 mm diameter droplet impacting a collector electrode in comparison with model from equation (1) for a calculated droplet charge of $10^8$ electrons.

With a regular stream of droplets, in order to measure the charge on individual droplets as they strike the collector, the electrode radius must fulfil the condition $R_c \leq X_{D-D}/3$, where $X_{D-D}$ is the distance between droplets. For a measured droplet rate of ~5 x 10$^4$ s$^{-1}$, and an average gas velocity of 16 m s$^{-1}$, the spacing between droplets is of the order of 320 μm and the limit is R$_C$ ≤ 100 μm.

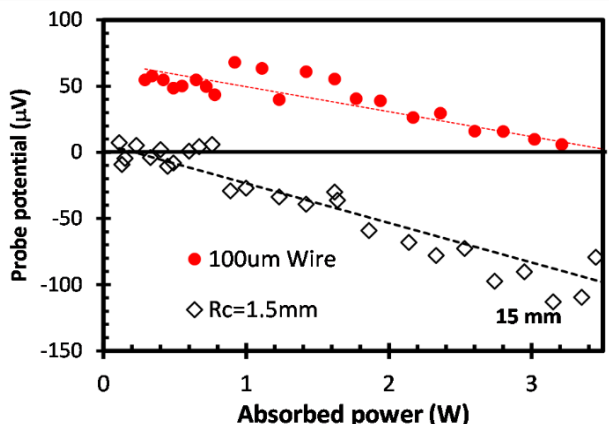

**Figure S2: Response of a narrow wire collector (100 mm diameter) compared to a large collector (3 mm diameter), to a plasma-exposed microdroplet stream. In the former, the collector potential remains positive indicating the absence of a significant number of droplet impacts compared to the large collector.**

Use of a 100 μm diameter wire collector proved unsuccessful; observed droplet current pulses (into 1 MΩ) were rare and the collector potential remained positive, as indicated in Figure S2. This is likely due to the majority of droplets being carried on gas streamlines around the thin wire, avoiding impact. As the single droplet condition (X$_{dd}$ >> 3R$_c$) cannot be satisfied in practice, the measured electrical response therefore corresponds to the superposition of a number of droplet charges present within the electrode response zone at any given instant, at different distances from the collector. The estimated total number of droplets in the plasma region, at any instant, is ~ 6.

## 1.1 Measurement Equivalent Circuit

The equivalent circuit of the measurement setup is shown in Figure S3 and shows three current sources between plasma lower (earth) electrode and the collector electrode. The first, i$_1$, is due to plasma ion and electron currents reaching the collector when the collector is located within the plasma or afterglow regions where charge densities are high. Since electrons are much more mobile than ions, the latter current dominates and the current direction is towards the plasma. For a collector situated well away from the plasma afterglow region, the majority of electrons have recombined in the gap before reaching the collector and a remnant small positive ion current remains, directed towards the collector. Finally, when droplets are added, a stream of negative charges, carried by the droplets, arrives at the collector forming a current, i$_3$, given by equation (1), and directed towards the plasma. To determine i$_3$ from measurements, i$_1$ and i$_2$ must be taken into account, i.e. i$_3$ = i$_{meas}$ – (i$_2$ – i$_1$). Accurate estimation of i$_1$ is not possible and it is therefore necessary to maintain a sufficient gap between plasma and collector to ensure i$_1$ ≈ 0. Under such conditions, and without droplets i$_2$ = i$_{meas}$, provided the introduction of droplets does not significantly affect plasma conditions.

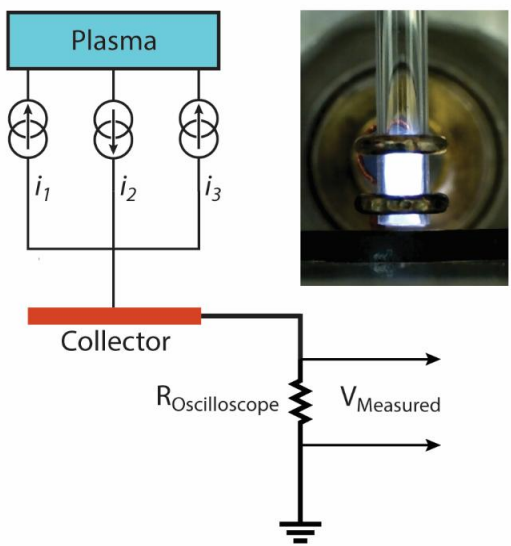

**Figure S3. Measurement equivalent circuit**

## 1.2 Plasma – collector gap

Ideally, to determine droplet charging within the plasma, the collector needs to be located as close as possible to the plasma, to limit surface charge recombination with the environment, while maintaining $i_1 \approx 0$. The afterglow region corresponds to a weak plasma with a carrier density diminishing with distance. The precise electrical extent of the afterglow, however, is not known. Langmuir probe spatial measurements in a similar system, using miniature (40 μm diameter probes), indicate a sharp rise in resistance just beyond (< 1 mm) the luminous region. Therefore, the afterglow region is assumed to correspond approximately with the visible luminous region beyond the electrodes, Figure S4 (a), and its length is observed to increase with absorbed power. However, with the addition of microdroplets, the afterglow length is much shorter, Figure S4 (b). To determine the effective electrical length of the afterglow against absorbed power, with and without droplets, current was measured using a thin wire (< 100 mm diameter) collector. The small collector area limits the number of droplet impacts and hence droplet current, $i_3 \approx 0$. Under these conditions, for a given gap, the current is positive when the power is sufficiently low and no contact exists between collector and afterglow. Once contact is made the current is progressively reduced as power increases, until with full contact, a negative current is measured.

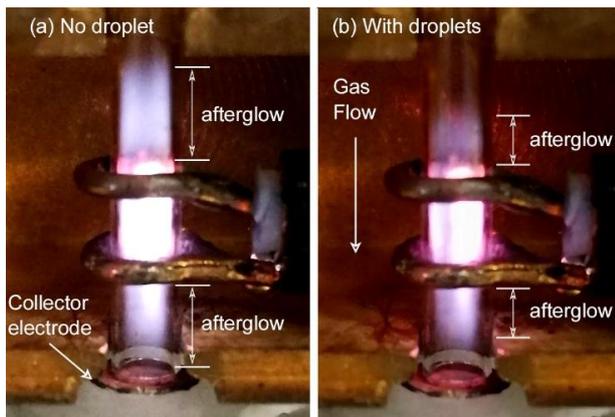

**Figure S4 (a) without droplets. The length of the luminous afterglow is illustrated by the arrow and compared with (b) when microdroplets are introduced.**

The average collector potential versus power, measured using a thin (< 100 mm) wire collector is shown in Figure S5, for plasma – collector gaps of 3 mm and 15 mm. Oscilloscope time-series traces show almost zero droplet pulses. At low powers the potential remains positive, indicated by the shaded regions. At higher powers, the afterglow length increases and when the collector is at least partially situated within the afterglow region, it then assumes a negative floating potential, as expected of any probe inserted into a plasma and commensurate with the local electron temperature. For 3 W power and at a distance of 15 mm, the collector potential is - 800 mV for the plasma without droplets but ~ 0 mV when droplets are introduced, Figure S5 (b). At short distances (3 mm), there is insufficient reduction in afterglow length when droplets are added and the potential – power characteristics are similar with and without microdroplets.

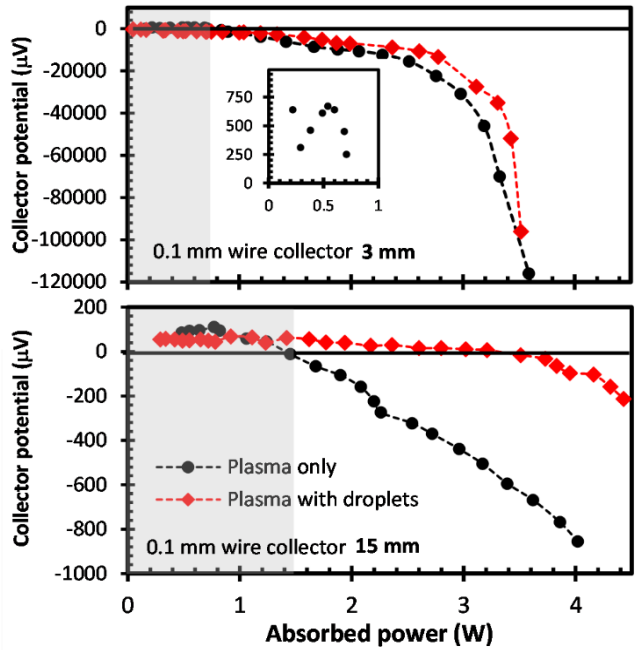

**Figure S5.** Comparison with and without droplets of collector potential ($V_C$) versus absorbed power for gaps of 3 mm and 15 mm from the plasma electrode, assuming droplet current $i_3 \approx 0$ due to the small size of the wire (100 mm diameter) electrode. The shaded regions indicate $V_C > 0$. Contact between electrode and plasma afterglow is indicated when $V_C < 0$. Inset (3 mm) shows shaded region positive potential values (plasma only).

### 1.3 Plasma ion current

Current measurements were taken in *uncoupled* mode i.e. at absorbed power levels low enough avoid contact between afterglow and collector, therefore ensuring $i_1 \approx 0$ and $i_{meas} = i_2 + i_3$. We obtain an estimate of $i_2$ from the plasma only measurements. For each value of plasma – collector gap, we obtain a linear fit to the current versus power characteristic up to the power threshold, Figure S6 (shaded region). To obtain the droplet induced current, $i_3$ (= $i_{meas} - i_2$), we obtain a linear fit to $i_{meas}$ versus power for each value of gap and subtract the $i_2$ fit, Figure S7. At the shortest gap, 3 mm, the maximum power before contact between afterglow and collector is 0.6 W. In order to estimate the maximum droplet charge within the plasma at the highest absorbed power, linear fits to $i_2$ and $i_{meas}$ are extrapolated to indicate the expected collector voltage at 5W, Figure S 8.

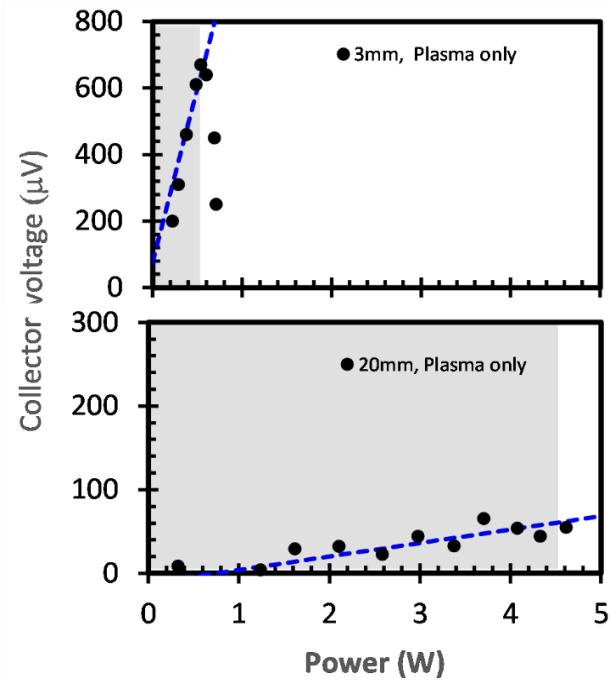

**Figure S6.** Positive ion current, $i_2$, versus power and associated linear fits for plasma – collector gaps of 3mm and 20 mm. The grey shaded region indicates where ion current is positive and the maximum power beyond which afterglow contact with collector occurs.

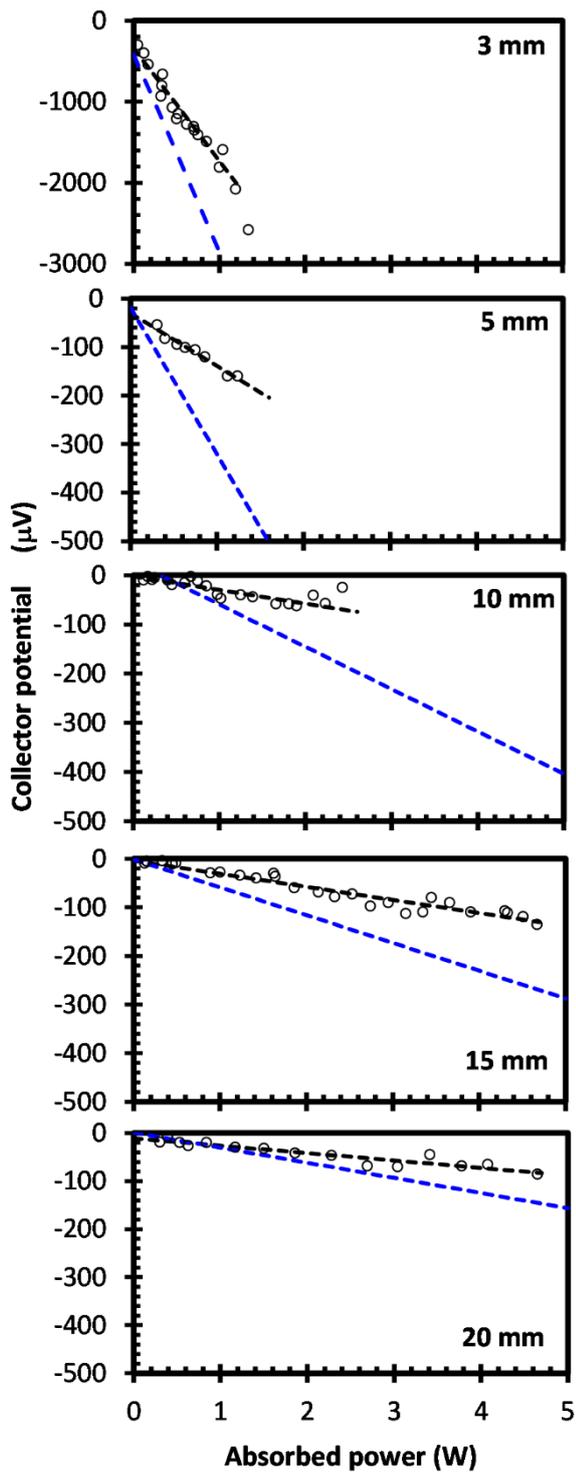

**Figure S7. Measured current, i_meas, versus power and associated linear fits (black). The blue line denotes the droplet induced current (i_meas – i_2) obtained by subtracting i_2 fit from i_meas fit at each distance.**

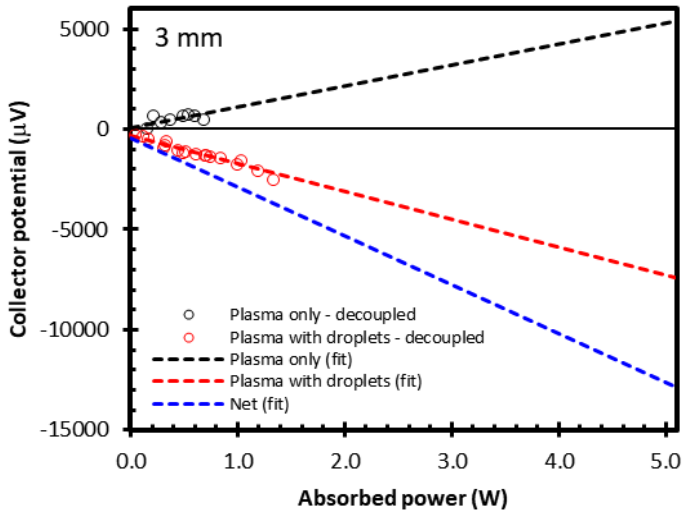

**Figure S 8:** Collector potential versus absorbed power at 3 mm. Least square fits ($R^2 > 0.94$) to uncoupled data, for both plasma only and plasma with droplets, enabled extrapolation to maximum power, 5W.

## Section S2. Current to charge conversion

To determine the average droplet charge, the average droplet current, $i_{drop}$ (= $i_{meas} - i_{ion}$), is compared with a simulation of the current equation (1). The simulated droplet stream contains 5 x $10^4$ droplets with diameters randomly selected from a lognormal probability distribution (CMD = 13.6 mm, GSD = 1.7) such that the diameter distribution is equivalent to that measured. A value of charge is allocated to each droplet according to its diameter and the diameter - charge relationship is estimated from charge transport simulation using electron density, electron temperature, positive ion and electron mobilities as input parameters – see section S5 Particle charge simulation in this document. The current to charge conversion factor was found to be 1.1 $10^{-7}$ V electron$^{-1}$ (±0.03%) for velocities 1 m s$^{-1}$ to 30 m s$^{-1}$, $T_e$ = 2 – 10 eV, $n_e$ = $10^{19}$ m$^{-3}$ – $10^{20}$ m$^{-3}$ and positive ion mobility values in the range 0.1 x $10^{-4}$ m$^2$ V$^{-1}$ s$^{-1}$ to 25 x $10^{-4}$ m$^2$ V$^{-1}$ s$^{-1}$, representing heavy water cluster ions to $He_2^+$.

# Section S3. Particle charge simulation

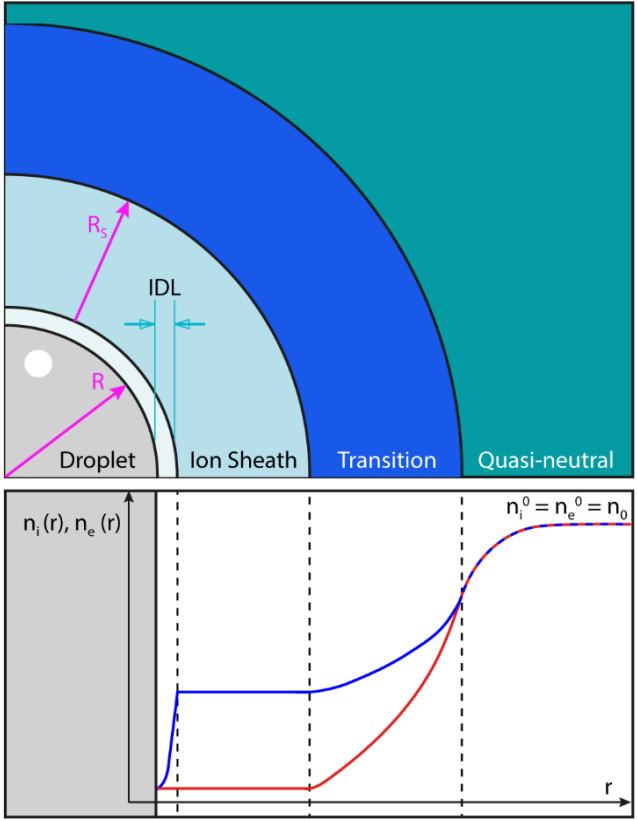

**Figure S 9: Spatial charge density around a droplet in a fully collisional plasma.**

Theoretical investigations into the charging of isolated particles (*probes*) within a collisional plasma were initiated in the 1960's and over the subsequent four decades, models were expanded and refined to encompass the full range of plasma regimes[2]. This culminated in the work of Patacchini and Hutchinson who provide the most comprehensive simulation and analysis of particle charging in low temperature collisional plasmas, handling sixteen different plasma and particle regimes determined by pressure, electron and ion temperatures, plasma density and particle radius[3]. Despite the extensive literature, charging models for fully collisional plasmas have yet to be validated experimentally. In deriving solutions to the continuum problem, the environment around the charged particle was divided into four regions, Figure S 9. Moving outwards from the particle, the ion diffusion boundary layer (IDL) at the particle surface becomes an ion sheath region (ISR), followed by a transition region (ITR) between sheath and the quasi-neutral (QNR) plasma region far from the particle. From the QNR – ITR interface outwards, the plasma density ($n_i(r) = n_e(r)$) increases up to that of the bulk plasma, $n_e^0 = n_i^0 = n_0$, far from the particle. The density profile, along with the electric field penetration into QNR, depends on the ion to electron temperature ratio, $T_i/T_e$, and the function $\mu_i I_e/\mu_e I_i$ where $\mu$ is the charged species mobility. In ITR, $n_i(r) \neq n_e(r)$ and in ISR, $n_i(r) > 0$, $n_e(r) = 0$. The sheath ion density then falls to zero over the narrow IDL region.

At atmospheric pressure, ion and electron mean free paths are much shorter than the electron Debye length and the droplet dimensions. As such, charge transport is governed by mobility – diffusion equations and for each charged species, the charge flux is given by

$$\Gamma_i = D_i \nabla n_i(r) - \mu_i n_i(r) \nabla \Phi(r) \tag{2}$$

$$\Gamma_e = D_e \nabla n_e(r) - \mu_e n_e(r) \nabla \Phi(r) \tag{3}$$

where n(r), D and μ are the charge density, diffusivity and mobility respectively, while Φ(r) is the self-consistent electrostatic potential on the droplet, given by Poisson's equation,

$$\nabla^2 \Phi(r) = -\frac{n_i(r) - n_e(r)}{\varepsilon_0} \tag{4}$$

Following the original approach of Su and Lam we consider a one-dimensional radial symmetric simulation scheme with spherical geometry[2]. Equations (2) - (4) are solved using COMSOL (v6.3) finite element software, assuming the following boundary conditions

  i.   $n_i(r) = n_e(r) = n_o$ at $r \gg R$ (QNR region)
  ii.  $F(r) = 0$ at $r \gg R$ (QNR region)
  iii. $n_i(r) = n_e(r) = 0$ at $r = R$ (Droplet surface)
  iv.  $F(r) = F_s$ at $r = R$ (Droplet surface)
  v.   $G_i(r) = G_e(r)$ at $r = R$ (Droplet surface)

and that mobility and diffusion coefficients obey Einstein's relation, i.e. $D_{i,e} = \mu_{i,e} k T_{i,e}/e$. Over 150,000 mesh elements were generated for a 1D spatial range extending up to 20 x R and equivalent to a resolution of 0.36 nm close to the surface.

## 3.1 Ion mobility

|   | Dominant ion(s) | Condition | Impurities | Mobility x$10^{-4}$ (m$^2$ V$^{-1}$ s$^{-1}$) | Reference |
|---|---|---|---|---|---|
| 1 | $He_2^+$ | High purity | 0 ppm $H_2O_v$, 0 ppm $N_2$, $O_2$ | 17 - 25 | 4,5 |
| 2 | $H_2O^+$, $H^+$ | Trace $H_2O$ vapour | < 10 ppm $H_2O_v$, 0 ppm $N_2$, $O_2$ | 21 | 6 |
| 3 | $(H_2O)_nH^+$ n = 5, 4 | Evaporation vapour | 1000 ppm $H_2O_v$, 0 ppm $N_2$, $O_2$ | 13 - 17 | 6 |
| 4 | $He_2^+$ | Dry air (trace quantity) | ≤ 1 ppm $H_2O_v$, ≤ 5 ppm $N_2$, $O_2$ | 17 - 25 | 4,5 |
| 5 | $NO^+$, $N^+$, $O_2^+$, $N_2^+$ | Dry air | ≤ 1 ppm $H_2O_v$, ≥ 10 ppm $N_2$, $O_2$ | 14 - 19 | 7–9 |
| 6 | $NO^+$, $(H_2O)_nH^+$ n = 5, 4 | Evaporation vapour, with air | ≥ 1000 ppm $H_2O_v$, ≥ 1000 ppm $N_2$, $O_2$ | 13 - 17 | 6,7 |
| 7 | $(H_2O)_nH^+$ n = 4 - 7 | Water cluster ions in saturated vapour halo |  | 0.68 | 10 |
| 8 | $(H_2O)_nH^+$ n > 7 | Large water cluster ions in saturated vapour halo |  | 0.1 | 11 |

**Table S1: Mobility values for dominant ions under various purity conditions. The carrier gas is He (99.999%).**

Plasma simulation (0D) using ZDplaskin software provided estimates of the dominant ion densities under various conditions which included pure He, He with $H_2O$ vapour from droplet evaporation and trace air impurity ingress[12,13]. The relevant outcomes are given in Table S1 along with the associated ion mobility values. At high purity $He_2^+$ dominates over $He^+$. With air ingress, > 10 ppm, the impurity ions dominate and at vapour concentrations > 1000 ppm (without air), water cluster ions were the dominant positive ion with cluster size increasing as concentration increased. Water cluster ions $(H_2O)_nH^+$ with n up to 55 (~1000 amu) have been observed by mass spectrometry for a He plasma jet into air[14]. Electron mobility values, $\mu_e$, were calculated using a Boltzmann solver and available cross-sections for He with various concentrations of $H_2O$ vapour up to 10,000 ppm, to evaluate possible conditions due to the presence of droplets[15,16]. In the simulation model, we assume constant values of $\mu_{i,e}$ and $D_{i,e}$ independent of electric field. We investigated the potential impact of electric field surrounding the droplet on mobility and resultant Q vs R characteristics. The maximum average electric field value, across the ion sheath region (r = 0 to r = R) was 2.5 x $10^6$ V m$^{-1}$. The electron mobility variation with electric field was determined using a Boltzmann solver[15,16]. The value varies by a maximum of 14%, from 0.102 m$^2$ V$^{-1}$ s$^{-1}$ at $T_e$ ~2 eV (E/N = 4 Td) to 0.116 m$^2$ V$^{-1}$ s$^{-1}$ at $T_e$ ~13 eV (E/N = 100 Td, i.e. E ~2.5 x $10^6$ V m$^{-1}$). Ion mobilities ($He_2^+$, $N_2^+$, $O_2^+$, $NO^+$) peak in the range 0 – 100 Td, i.e. 0 - 2.5 x $10^6$ V m$^{-1}$, with a maximum deviation of ~ 9% ($He_2^+$). For water cluster ions the deviation is ~ 2%. The impact of mobility variation on estimated charge was negligible.

## 3.2 Simulated Q vs R

For a range of plasma density, $n_o$, values from $10^{18}$ m$^{-3}$ to $10^{21}$ m$^{-3}$, droplet radius values from 3 μm to 25 μm and $T_e$ values from 1 eV to 13 eV, the droplet floating potential, $\Phi_s$, was iterated until boundary condition (v) was met. This was repeated for a range of ion mobility values, from 0.1 x $10^{-4}$ m$^2$ V$^{-1}$ s$^{-1}$ to 25 x $10^{-4}$ m$^2$ V$^{-1}$ s$^{-1}$ to account for possible variation in plasma ion chemistry, Table S1. Plots of droplet charge against radius are shown in Figure S10, for ion mobility values of 0.1 x 10$^{-4}$ m$^2$ V$^{-1}$ s$^{-1}$ (water cluster ions) to 25 x 10$^{-4}$ m$^2$ V$^{-1}$ s$^{-1}$ (He$_2^+$ ions). A reduction in ion mobility values leads to an increase in the acquired charge.

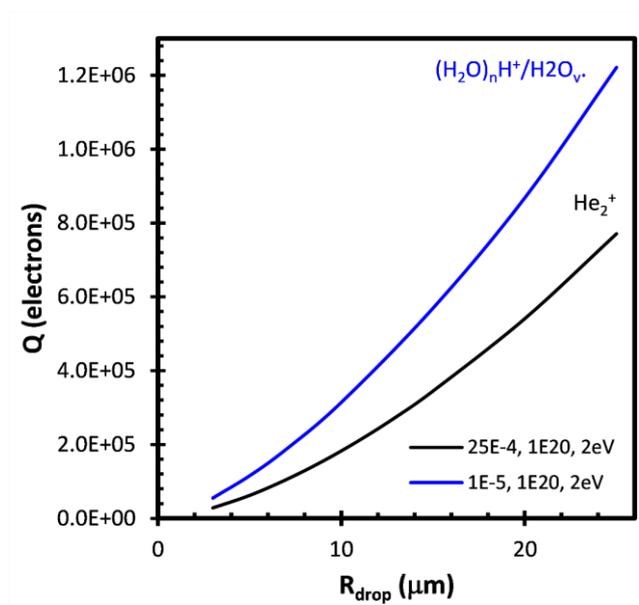

**Figure S10: Droplet charge vs radius for $n_e = 10^{20}$ m$^{-3}$, $T_e = 2$ eV and positive ion mobilities of 0.1 x 10$^{-4}$ m$^2$ V$^{-1}$ s$^{-1}$ (water cluster ions) and 25 x 10$^{-4}$ m$^2$ V$^{-1}$ s$^{-1}$ (He$_2^+$ ions).**

### 3.3 Measured charge – radius (Q – R) relationship

We measure an average charge value on the droplets from a droplet stream with a known size distribution. Given the relationship between acquired charge and droplet radius observed from simulation, Figure S10, it is possible to estimate an equivalent charge distribution from the droplet size distribution. We determine a least squares 2$^{nd}$ order polynomial fit to the Q – R characteristic obtained from simulation and a relative charge value is given to each droplet size according to the model fit. Then, using the measured size distribution, the charge on each droplet is scaled uniformly until the calculated distribution average charge is equal to that measured. The resultant plot of droplet charge distribution is shown in Figure S11, using an average charge value of 2.5 x 10$^5$ electrons obtained from measurements taken at high absorbed power.

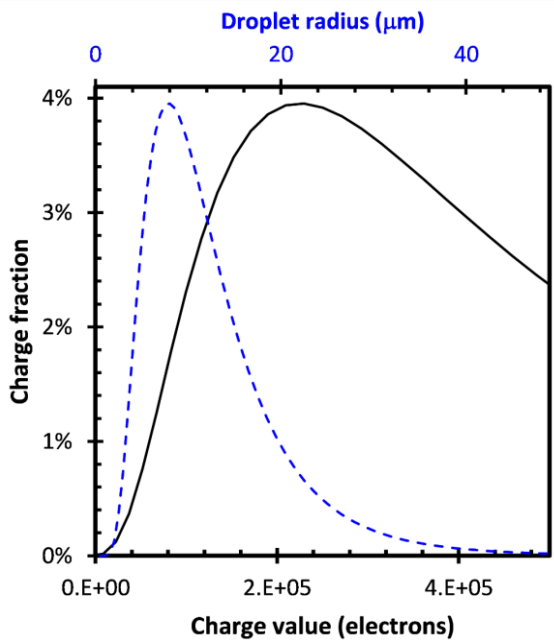

**Figure S11: Plot of equivalent charge fraction across the droplet distribution shown against radius, R, and charge value, Q. The relationship between Q and R was derived from simulation ($n_e = 10^{20}$ m$^{-3}$, $T_e = 2$ eV, $\mu_{ion} = 10^{-5}$ m$^2$ V$^{-1}$s$^{-1}$) and a 2$^{nd}$ order polynomial fit obtained with coefficients $a_2 = 1038$, $a_1 = 24163$, $a_0 = -30099$. The sum of distribution Q was scaled until it equalled the measured Q value of 2.5 x 10$^5$ electrons (5 W, 3 mm – extrapolated to 0 mm)**

# Section S4. Electron density and temperature

## 4.1 Electron temperature

We have previously measured the temperature (1.1 eV) in a similar plasma but ignited in argon rather than helium, using spectroscopic measurements and a collisional-radiative model[17,18]. A similar approach is not possible with He and instead an electron temperature value of 2.0 eV was estimated from plasma chemistry simulation, whereby the value was adjusted until the simulation output matched the measured chemical species ($H_2O_2$, $OH^{\bullet}$)[13]. This is in agreement with that predicted by theoretical analysis which suggests $T_e$ in He is approximately double that in Ar, under the same conditions, due to the higher energy of the 1st excited state in He[19,20].

## 4.2 Electron density

Since He is much lighter than Ar, greater diffusional losses occur and the electron density is generally lower for similar operating conditions and typically below the measurement capability of standard spectroscopic quantification techniques. We therefore obtained an estimate of $n_e$ from inline close-coupled impedance measurements, assuming a simple cylindrical plasma geometry between the electrodes, Figure S 12. Circuit analysis with and without the plasma ignited provided values for plasma resistance and sheath capacitance (along with air and quartz parasitic impedances) from which plasma sheath width, $X_s$, can be extracted. With the area given by $\pi(D - 2X_s)^2/4$, where D is the plasma diameter, $n_e$ is obtained from

$$R_{plasma} = \frac{L_{plasma}}{e \mu n_e A} \quad (5)$$

The electron mobility values, $\mu(T_e)$, were calculated using a Boltzmann solver and available cross-sections for He with various concentrations of $H_2O$ vapour up to 10,000 ppm, to evaluate possible variation due to the presence of droplets[15,16]. In Figure S 13, the calculated variation in plasma density without and with droplets is shown for $T_e \sim 2$ eV and a vapour concentration of 500 ppm. With the inclusion of water vapour up to 10,000 ppm (not shown), calculated $n_e$ values deviated by < 1% from those at 500 ppm $H_2O$, while varying $T_e$ from 1 eV to 5 eV leads to an increase in $n_e$, as indicated by the vertical bars.

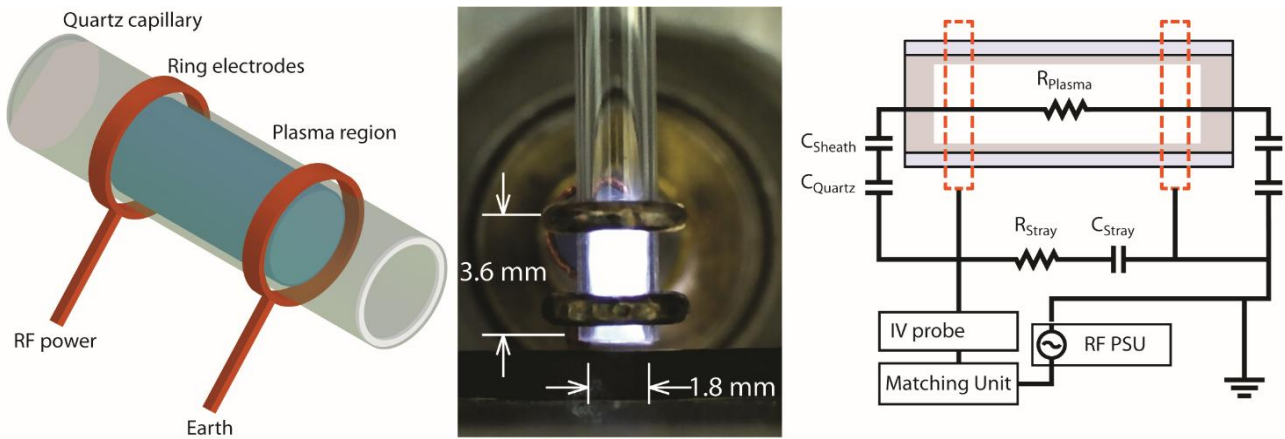

**Figure S 12: Quartz capillary tube with coaxial plasma electrodes and equivalent impedance circuit.**

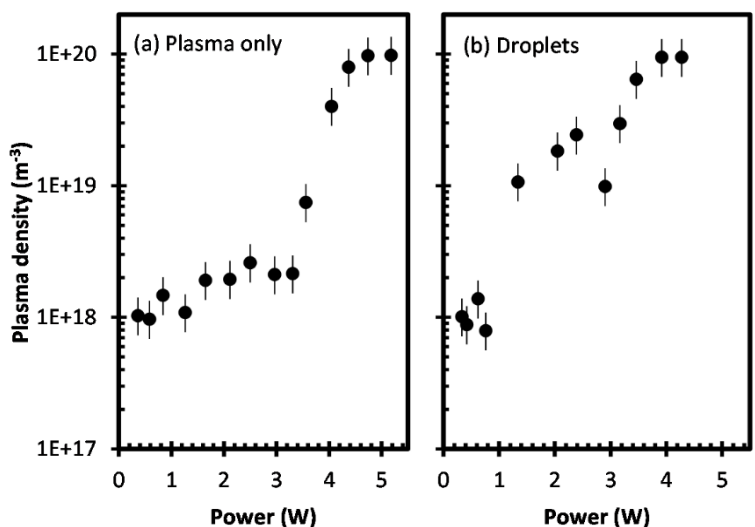

Figure S 13: Calculated values of plasma density versus absorbed power from an impedance – mobility model without (a) and with (b) droplets for an electron temperature of 2 eV. The vertical lines indicate the variation in $n_e$ due to varying electron temperature from 1 eV to 5 eV.

## Section S5. $H_2O_2$ measurement

$H_2O_2$ concentration was calculated from UV absorbance spectroscopy measurements of titanium (IV) oxysulphate (TiS) solution (Sigma Alrich 495 379). A deuterium lamp (Ocean Optics DH-2000-BAL) was used in combination with an Ocean Optics spectrometer (QE65 Pro). TiS was added to the sample cuvette at a concentration of 100 mM. It then reacts with $H_2O_2$ to produce pertitanic acid with a characteristic absorbance at 407 nm. Absolute calibration was performed using the absorbance of several samples containing known $H_2O_2$ concentrations. 300 mL of plasma exposed droplet liquid was collected in 3700 mL of $H_2O$. The sample collection time was 30 min, Figure S 14. With a dilution factor of 0.075, the peak $H_2O_2$ concentration in the droplet was calculated to be 33 mM. For an estimated flight time of 120 µs in the plasma, the calculated $H_2O_2$ generation rate was 275 M s$^{-1}$. The collection liquid was exposed to plasma effluent without droplets for the same sample collection time. The measured concentration was approximately constant at 177 µM.

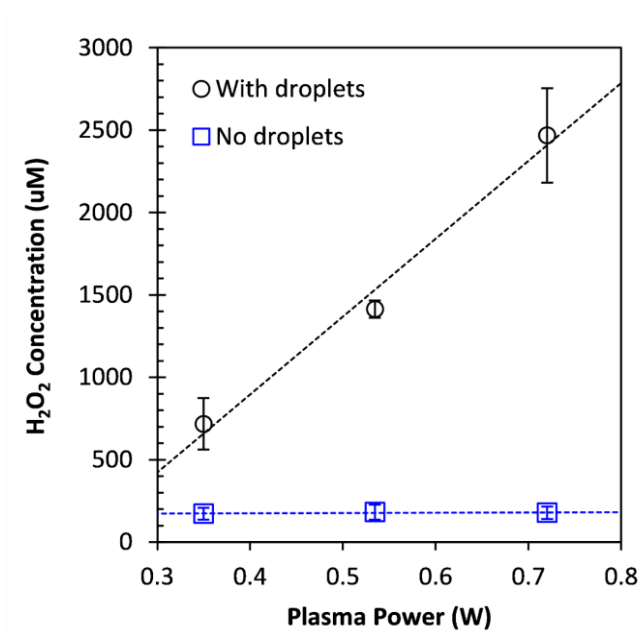

Figure S 14: Measured $H_2O_2$ concentration in collection liquid downstream from plasma (~100 mm), with and without droplets. The actual droplet concentration is calculated using a dilution factor of 0.075.

## Section S6. References